\documentclass[aps, prd, groupedaddress, preprint, tightenlines,  eqsecnum, nofootinbib, showpacs]{revtex4}
\usepackage{epsfig}
\usepackage{setspace}
\usepackage{graphicx}
\usepackage{leqno}
\usepackage{cancel}
\usepackage{amsmath}
\usepackage{amsfonts}
\usepackage{amssymb}
\usepackage{enumerate}

\usepackage{listings}

\usepackage{axodraw, float, slashed, graphicx, amssymb, amsmath}
\usepackage[usenames, dvipsnames]{xcolor}
\usepackage{booktabs}

\usepackage[margin=2.2cm, a4paper, includefoot]{geometry}

\def\equn{\refstepcounter{equation}\eqno({\rm \theequation})}
\def\Li{\mathop{\hbox{\rm Li}}\nolimits}

\def\ceta{q}

\def\spa#1.#2{\left\langle#1\,#2\right\rangle}
\def\spb#1.#2{\left[#1\,#2\right]}
\def\la{\langle}
\def\ra{\rangle}

\newcommand{\tree}{\text{tree}}

\DeclareMathOperator{\tr}{\mathrm{tr}}

\newcommand\figref[1]{fig.~\ref{#1}}
\def\eps{\epsilon}

\def\be{\begin{equation}}
\def\ee{\end{equation}}

\def\Log{{\rm Log}}
\begin{document}

\hfill\today

\title{Two-loop five-point all plus helicity Yang-Mills amplitude}

\author{David~C.~Dunbar and Warren~B.~Perkins}

\affiliation{
College of Science, \\
Swansea University, \\
Swansea, SA2 8PP, UK\\
\today
}

\begin{abstract}

We re-compute the recently derived two-loop five-point all plus Yang-Mills amplitude using Unitarity and Recursion. 
Recursion requires augmented recursion to determine the sub-leading pole.  Using these methods the simplicity of this amplitude is understood. 

\end{abstract}

\pacs{04.65.+e}

\maketitle

\section{Introduction}

Computing perturbative scattering amplitudes is a key challenge in Quantum Field theory both for comparing theories with experiment and for
understanding the symmetries and consistency of theories. 
Explicit analytic expressions for scattering amplitudes have proved to be
useful windows into the behaviour of the underlying theory.   Technical developments have been crucial to computing these amplitudes. 
Two key methods  based upon unitarity~\cite{Bern:1994zx,Bern:1994cg} and on-shell recursion~\cite{Britto:2005fq} have produced a great many 
spectacular results particularly for maximally supersymmetric field theories.

Recently the two-loop all-plus five-point amplitude has been computed in QCD~\cite{Badger:2013gxa,Badger:2015lda} using $d$-dimensional unitarity techniques. 
Subsequently this amplitude was presented in a very elegant and compact form~\cite{Gehrmann:2015bfy}.  In this form the amplitude consists of a piece 
driven by the Infra-Red (IR) structure of the amplitude and a ``remainder'' piece. In this article we 
demonstrate how this form can be generated using a combination of four-dimensional unitarity and (augmented) 
recursion which provides an understanding of the simplicity of the amplitude.

Following Gehrmann et al.~\cite{Gehrmann:2015bfy}, the all-plus amplitude at leading colour may be written\footnote{The factor $c_{\Gamma}$
is defined as $\Gamma(1+\epsilon)\Gamma^2(1-\epsilon)/\Gamma(1-2\epsilon)/(4\pi)^{2-\epsilon}$. Note this gives a factor of $1/(16\pi^2)$ relative to other normalisations in the literature.}
\begin{align}
{\mathcal A}_{5}(1^+, 2^+, 3^+, 4^+, 5^+) |_{\rm leading \; color}=& g^3  \sum_{L \ge 1} \left( g^2 N c_{\Gamma}\right)^L  
\times \sum_{\sigma \in S_{5}/Z_{5}} {\rm tr}(T^{a_{\sigma(1)}} T^{a_{\sigma(2)}} T^{a_{\sigma(3)}} T^{a_{\sigma(4)}} T^{a_{\sigma(5)}}) 
\notag\\ &
\times A^{(L)}_{5}(\sigma(1)^{+},  \sigma(2)^{+} ,\sigma(3)^{+} ,\sigma(4)^{+}, \sigma(5)^{+})\,
\end{align}
and the object we wish to compute is the color-stripped two-loop amplitude $A^{(2)}_5(1^+,2^+,3^+,4^+,5^+)$.

The IR and UV behaviour of the amplitude are well specified~\cite{Catani:1998bh} and motivate a partition of the amplitude:
\begin{align}\label{definitionremainder}
A^{(2)}_{5} =& A^{(1)}_{5} \left[ - \sum_{i=1}^{5} \frac{1}{\epsilon^2} \left(\frac{\mu^2}{-s_{i,i+1}}\right)^{\epsilon} 
+\frac{5\pi^2}{12} \right] +  \;F^{(2)}_{5}  + {\mathcal O}(\eps)\, .
\end{align}
The leading term in eq.(\ref{definitionremainder}) contains the necessary IR and UV terms. In this equation $A^{(1)}_{5}$ is the all-$\epsilon$ form of 
the one-loop amplitude. 
The remainder function $F_5^{(2)}$ is to be determined.  We further organise $F_5^{(2)}$ into cut-constructible and rational pieces,
\begin{equation}
F_5^{(2)} = F_5^{cc}+R_5^{(2)}\; .
\end{equation}

\section{Cut Constructible Pieces} 

In \cite{Badger:2013gxa} $d$-dimensional unitarity was used to compute a master integral representation of the full two-loop five-point all-plus amplitude 
$A^{(2)}_5(1^+,2^+,3^+,4^+,5^+)$. When using $d$-dimensional unitarity the cuts of the amplitude have cut legs  defined in $d=4-2\epsilon$ dimensions. 
Given a Feynman diagram expansion of an amplitude, 
polynomial reduction~\cite{Mastrolia:2011pr,Badger:2012dp,Zhang:2012ce,Mastrolia:2012an,Badger:2012dv,Mastrolia:2012wf,Mastrolia:2013kca} 
can be used to obtain a corresponding set of master integrals. The reduction process involves cutting each diagram
and repeatedly isolating the irreducible contribution on each cut. For example, the pentabox diagram 
has all eight propagators in loops and has a non-vanishing 
eight-fold cut. The first step of the division is to evaluate the numerator on the eight-fold cut, thus determining the non-vanishing contribution when all 
eight propagators vanish. The remainder is then evaluated
on all possible seven-fold cuts and so on. This approach can also be used in a similar manner to the one-loop unitarity method. 
Each set of cuts determines a partition of the full set of Feynman diagrams 
into blocks which must be of lower loop order, in this case tree or one-loop blocks. Summing over all diagrams yields an on-shell amplitude for each block. 
The contribution from each cut is then determined using
the product of these amplitudes for each block.

\begin{figure}[H]
\centerline{
    \begin{picture}(170,150)(-150,-20)    
     \Line( 0, 0)( 0,60)
     \Line( 0,60)(60,60)
     \Line(60,60)(60, 0)
     \Line(60, 0)( 0, 0)
     \Line( 0, 0)(-15,-15)
     \Line( 0,60)(-15,75)
     \Line(60,60)(60,75)
     \Line(60,60)(75,60)
     \Line(60, 0)(68,-8)
   \CCirc(60,60){5}{Black}{Purple} 
    \end{picture} 
    \begin{picture}(170,150)(-90,-20)    
     \Line( 0, 0)(30,60)
     \Line(60, 0)(30,60)
     \Line(60, 0)( 0, 0)
     \Line( 0, 0)(-10,0)
     \Line( 0, 0)(0,-10)
     \Line(30,60)(45,75)
     \Line(30,60)(15,75)
     \Line(60, 0)(68,-8)
    \CCirc(30,60){5}{Black}{Purple} 
    \end{picture} 
   \begin{picture}(170,150)(-40,-20)    
     \Line( 0, 0)(30,60)
     \Line(60, 0)(30,60)
     \Line(60, 0)( 0, 0)
     \Line( 0, 0)(-8,-8)
     \Line(15,60)(45,60)
     \Line(30,60)(30,75)
     \Line(60, 0)(68,-8)
   \CCirc(30,60){5}{Black}{Purple} 
    \end{picture} 
    \begin{picture}(170,150)( 30,-20)  
     \CArc(30,30)(20,0,360)
     \Line(10,30)( 2,38)
     \Line(10,30)( 2,22)
     \Line(50,30)(65,38)
     \Line(50,30)(65,22)
     \DashLine(50,30)(63,30){2}
     \DashLine(10,30)(-3,30){2}
    \CCirc(50,30){5}{Black}{Purple}    
 \end{picture} 
    }
    \caption{Contributions to the two-loop amplitudes involving an all-plus loop (indicated by $\bullet$)}
    \label{fig:oneloopstyle}
\end{figure}
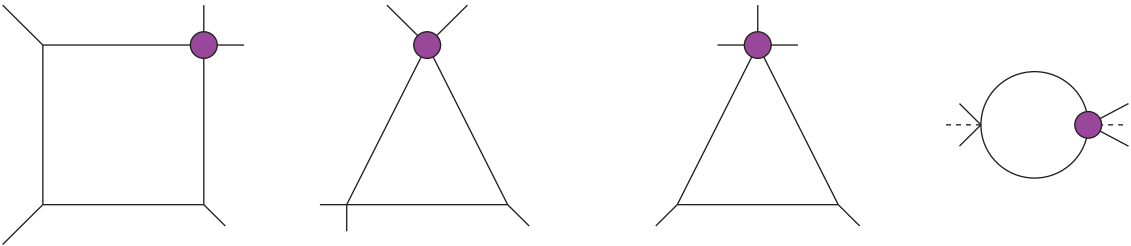

Here, alternatively, four-dimensional amplitudes will be used to determine the cut-constructible pieces of the
remainder function and then the  remaining rational pieces will be calculated recursively. 
For the all-plus amplitude considerable simplification arises when we restrict ourselves to four-dimensional cuts because all 
four-dimensional cuts of the one-loop all-plus amplitude vanish. After discarding scale free cuts, the reduction process only receives contributions from 
structures of the forms shown in \figref{fig:oneloopstyle}, where the
$\bullet$ denotes an un-cut one-loop all-plus amplitude.
These contributions involving the all-plus one-loop amplitude can be evaluated using one-loop techniques with the one-loop sub-amplitude as a vertex.
The $n$-point all-plus one-loop amplitude is~\cite{Bern:1993qk}
\begin{align}
A^{(1)}(1^+,2^+,\cdots,n^+)=-{i\over 3}\sum_{1\leq k_1<k_2<k_3<k_4\leq n} 
{\spa{k_1}.{k_2} \spb{k_2}.{k_3}\spa{k_3}.{k_4}\spb{k_4}.{k_1} \over \spa{1}.2\spa{2}.3 \cdots\spa{n}.1}  
+O(\epsilon) \; . 
\end{align}
Note that for the four-point amplitude there are no  box functions with non-vanishing coefficients and the remainder function for the four-point amplitude 
is purely rational~\cite{Bern:2002tk}.

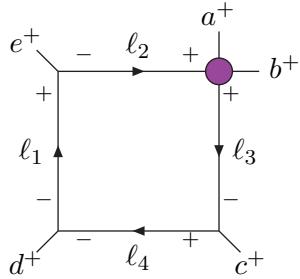
\begin{figure}[H]
\centerline{
    \begin{picture}(170,150)(-80,-20)    
     \ArrowLine( 0, 0)( 0,60)
     \ArrowLine( 0,60)(60,60)
     \ArrowLine(60,60)(60, 0)
     \ArrowLine(60, 0)( 0, 0)
     \Line( 0, 0)(-8,-8)
     \Line( 0,60)(-8,68)
     \Line(60,60)(60,75)
     \Line(60,60)(75,60)
     \Line(60, 0)(68,-8)
     \Text(-12,-12)[c]{$d^+$}  
     \Text(-12, 72)[c]{$e^+$}  
     \Text( 72,-12)[c]{$c^+$}  
     \Text( 85, 62)[c]{$b^+$}  
     \Text( 60, 82)[c]{$a^+$} 
     \Text( 60, 60)[c]{$\bullet$}
     \CCirc(60,60){5}{Black}{Purple}
     \Text(-10, 30)[c]{$\ell_1$}  
     \Text( 70, 30)[c]{$\ell_3$}  
     \Text( 30,-10)[c]{$\ell_4$}  
     \Text( 30, 70)[c]{$\ell_2$}
     \Text( 50, 65)[c]{$^+$}
     \Text( 10, 65)[c]{$^-$}     
     \Text( 50, -5)[c]{$^+$}
     \Text( 10, -5)[c]{$^-$}     
     \Text( 65, 50)[c]{$^+$}     
     \Text( 65, 10)[c]{$^-$}     
     \Text( -5, 50)[c]{$^+$}     
     \Text( -5, 10)[c]{$^-$}     
    \end{picture} 
    }
    \caption{The labelling and internal helicities of the quadruple cut.}
    \label{fig:oneloopbox}
\end{figure}

The box contribution is readily evaluated using a quadruple cut~\cite{BrittoUnitarity}. With the labelling of \figref{fig:oneloopbox} the cut momenta are
\begin{align}
\ell_1={\spa{c}.d\over\spa{e}.c} \bar\lambda_d \lambda_e
\,\, &,\,\,
\ell_2={\la c|P_{de}|\over\spa{e}.c}  \lambda_e
\,\, ,\,\, 
 \ell_3={\la e|P_{cd}|\over\spa{e}.c}  \lambda_c
\,\, ,\,\, 
 \ell_4={\spa{e}.d\over\spa{e}.c} \bar\lambda_d \lambda_c
 \,\, ,\,\, 
\end{align}
giving the coefficient of the box function\footnote{External legs attached to the one-loop corner are enclosed in brackets thus $\{ \cdots\}$} 
\begin{align}
{\cal C}_{\{a,b\},c,d,e} =& 
M_4^{(1)} (a^+,b^+,l_3^+,l_2^+) \times  M_3^{\tree}( l_3^-, c^+ ,l_4^+ ) \times  M_3^{\tree}(l_4^-, d^+, l_1^-)  \times M_3^{\tree}(l_1^+, e^+, l_2^-)
\notag \\
=&
{i\over 6}{\spb{a}.b^2\spb{c}.d\spb{d}.e\over \spa{c}.e } 
\; .  
\end{align}
This is the coefficient of the integral function $I_4^{\rm 1m}(s_{cd},s_{de},s_{ab})$ where~\cite{Bern:1993kr}
\def\rg{r_{\Gamma}}
\def\e{\epsilon}
\def\Li{{\rm Li}}
\begin{align}
I^{\rm 1m}_4(S,T,M^2) = -{2  \over S T }
\Biggl[&-{1\over\e^2} \Bigl[ (-S)^{-\e} +
(-T)^{-\e} - (-M^2)^{-\e} \Bigr] \cr
\notag \\ 
 & + \Li_2\left(1-{ M^2 \over S }\right)
   + \ \Li_2\left(1-{M^2 \over T}\right)
   +{1\over 2} \ln^2\left({ S  \over T}\right)
+\ {\pi^2\over6}
\Biggr] 
\end{align}
and overall factors of $c_{\Gamma}$ have been  removed according to the normalisation of eq.~(\ref{definitionremainder}).
This integral function splits into singular terms plus a remainder $I_4^{\rm 1m}=I_4^{\rm 1m:IR}+I_4^{\rm 1m:F}$
where
\begin{equation}
I_4^{\rm 1m:IR}(S,T,M^2) \equiv  -{2\over S T }
\Biggl[-{1\over\e^2} \Bigl[ (-S)^{-\e} +
(-T)^{-\e} - (-M^2)^{-\e} \Bigr] \Biggr]
\; . 
\end{equation}

The IR infinite terms, $I_4^{\rm 1m:IR}$, in this combine with the IR infinite terms in the triangle integral functions to produce the correct IR infinite terms
in the two-loop amplitude while the 
finite pieces, $I_4^{\rm 1m:F}$, contribute to the remainder function.

The triangle contributions can be evaluated using triple cuts~\cite{Bidder:2005ri,Darren,BjerrumBohr:2007vu,Mastrolia:2006ki}   
and a canonical basis~\cite{Dunbar:2009ax}.       
Each one-mass triangle $I_{3}^{1\rm m}(s_{ed})$ has two helicity configurations which give identical
coefficients,
\begin{align}
{\cal C}_{\{a,b,c\},d,e} ={i\over 6} { s_{de} \over \spa{e}.d \spa{a}.b \spa{b}.c}   
 \Biggl( &       
   s_{ba} \left( { \spb{e}.a  \over \spa{d}.c}-{\spb{d}.a  \over \spa{e}.c}\right) 
  -s_{bc} \left( { \spb{e}.c \over \spa{d}.a}-{ \spb{d}.c  \over \spa{e}.a}\right) 
-2\spb{d}.e \spb{a}.c         
\Biggr)     
\end{align}
and the integral function is
\begin{align}
I_{3}^{1\rm m}(K^2) = {1\over\e^2} (-K^2)^{-1-\e}.
\end{align}
Similarly the two mass triangle contributions are
\begin{align}
 {\cal C}_{\{a,b\},c,(d,e)}= {i\over 6} {\spb{a}.b^2\over \spa{c}.d\spa{d}.e\spa{e}.c}[c|P_{de}|c\ra  I_{3}^{2 \rm m}\bigl( s_{ab},s_{de}\bigr)
 \; , 
\end{align}
where the two-mass triangle function is,
\begin{align}
I_{3}^{2 \rm m}\bigl(K_1^2,K_2^2\bigr)= {1\over\e^2}
{(-K_1^2)^{-\e}-(-K_2^2)^{-\e} \over  (-K_1^2)-(-K_2^2) }\ .
\end{align}

The bubble contributions can be evaluated using double cuts and a canonical basis~\cite{Dunbar:2009ax}. The product of amplitudes 
in each double cut is order $\ell^{-2}$ and hence the bubble coefficients vanish.   This is consistent with the absence of $\eps^{-1}$ 
singularities in the amplitude.

The boxes, one-mass and two-mass triangles all have IR infinite terms of the form
$$
{1\over\e^2}(-K^2)^{-\e}
\; . 
$$
A specific choice of $K^2=s_{ab}$  arises from 
three box functions,   
$$
I_4^{\rm 1m}( \{a,b\},c,d,e)  :\;\; I_4^{\rm 1m}( \{c,d\},e,a,b)  :\;\;  I_4^{\rm 1m}( \{d,e\},a,b,c ) \; ,
$$
four two-mass triangle functions,
$$
I_3^{\rm 2m} ( \{a,b\},c, ( d,e) ) :\;\;
I_3^{\rm 2m} (\{a,b\}, ( c,d ),e ) :\;\;
I_3^{\rm 2m} (\{c,d\},e,( a,b ) ) :\;\;
I_3^{\rm 2m} (\{d,e\}, (a,b ),c )
$$
and a single one-mass triangle function $I_3^{\rm 1m}( \{c,d,e\},a,b)$.
Summation over the box and triangle 
contributions gives an
overall coefficient of $A^{(1),\epsilon^0}_5(a^+,b^+,c^+,d^+,e^+)$,
\begin{align}
&
\left(   \sum  {\cal C}_{\{a,b\},c,d,e} I_4^{1-mass}
+\sum  {\cal C}_{\{a,b,c\},d,e} I_{3}^{1\rm m}
+\sum  {\cal C}_{\{a,b\},c,(d,e)}  I_{3}^{2 \rm m}  \right)_{IR}
\notag \\ = &
A^{(1),\epsilon^0}_5(a^+,b^+,c^+,d^+,e^+)
\times \sum_{i=1}^{5} \frac{1}{\epsilon^2} \left(\frac{\mu^2}{-s_{i,i+1}}\right)^{\epsilon} ,
\end{align}
where $A^{(1),\epsilon^0}_5(a^+,b^+,c^+,d^+,e^+)$ is the order $\epsilon^0$ truncation of the one-loop amplitude.
A key step is to promote the coefficient of these terms to be the all-$\epsilon$ form of the one-loop amplitude which 
then gives the correct singular structure of the amplitude.

The finite part of the one-mass boxes, $I_4^{\rm 1m:F}$, then gives the cut-constructible part of the remainder function,  
\begin{equation}
F_5^{cc}=
\sum
{i\over 6}{\spb{a}.b^2\spb{c}.d\spb{d}.e\over \spa{c}.e }  
\times  \Biggl( 
 -{2 \over s_{cd} s_{de}  } \Biggr)
\Biggl[ 
\Li_2\left(1-{ s_{ab} \over s_{cd} }\right)
   +  \Li_2\left(1-{s_{ab}  \over s_{de} }\right)
 +{1\over 2} \ln^2\left(     {  s_{cd}  \over s_{de} }\right)
+ {\pi^2\over6}
\Biggr]  \; ,
\label{eq:cc}
\end{equation}
in agreement with ref.~\cite{Gehrmann:2015bfy}. 
This combination of dilogarithms can either be viewed as a truncated box or, as recognised in ref.~\cite{Gehrmann:2015bfy}, the $D=8$ dimensional box.  
This combination arises in one-loop amplitudes without $\eps^{-2}$ IR singularities~\cite{Bidder:2005ri,BrittoUnitarity}.

\section{Rational Pieces}

We obtain $R_5^{(2)}$ using the on-shell recursion techniques introduced by Britto-Cachazo-Feng and Witten (BCFW) to compute
tree amplitudes~\cite{Britto:2005fq}. 
In this technique  the amplitude is found by introducing a shift that transforms the amplitude into an analytic function of a complex 
parameter, $z$, then using Cauchy's theorem to reconstruct the rational part from its poles:
\begin{align}
\label{recstatement}
{1\over 2\pi i}\oint {A(z)\over z}= A(0) +\sum_{z_j\neq 0} {\rm Res}\Bigl[ {A(z)\over z}\Bigr]\Bigr\vert_{z_j}\; .
\end{align}
Taking the contour to be the circle at infinity, the left hand side of eq.(\ref{recstatement}) vanishes provided the shifted amplitude 
vanishes for large values of $z$.  As the poles in the amplitude are determined by its factorisations,
the unshifted amplitude is obtained in terms of lower point on-shell tree amplitudes:
\begin{equation}
A_n^\tree (0) \; = \; \sum_{i,\lambda} {A^{\tree,\lambda}_{r_i+1}(z_i)
  {i\over K^2}A^{\tree,-\lambda}_{n-r_i+1}(z_i)} \; .
\label{RecursionTree}
\end{equation}
The usual shift involves a pair of spinors:
\begin{equation}
\bar\lambda_{{a}}\to \bar\lambda_{\hat{a}} =\bar\lambda_a - z \bar\lambda_b 
\qquad , \qquad
\lambda_{{b}}\to\lambda_{\hat{b}} =\lambda_b + z \lambda_a 
. 
\label{BCFWshift}
\end{equation}

We wish to apply on-shell recursion to $R_5^{(2)}$ however there are some obstacles.  Firstly the shift of eq.~(\ref{BCFWshift})
does not produce an expression which has the correct cyclic symmetry. This is usually a signature that the expression does not vanish at infinity as
may also be inferred from the behaviour of the cut-constructible terms.  
(This can be checked {\it a posteriori} from the expressions in ref~\cite{Gehrmann:2015bfy}.)

Instead we use the shift~\cite{Risager:2005vk,BjerrumBohr:2005jr}
\begin{align}
\lambda_c\to &\lambda_{\hat c} = \lambda_c +z\spb{d}.e \lambda_\eta \,,
\notag \\
\lambda_d\to &\lambda_{\hat d} = \lambda_d +z\spb{e}.c \lambda_\eta \,,
\notag \\
\lambda_e\to &\lambda_{\hat e} = \lambda_e +z\spb{c}.d \lambda_\eta \,,
\label{KasperShift}
\end{align}
where $\lambda_\eta$ is an arbitrary spinor.  
Under this shift the cut-constructible terms vanish as $z\to \infty$, an indication that the rational part will also have well behaved asymptotics.

A further issue is the existence of double poles in the amplitude.  These arise beyond tree level.  In principle these are not a barrier 
to computation  since, if we have a function whose expansion about $z_i$ is 
$$
f(z) = {a_{-2} \over (z-z_i)^2 }+
{a_{-1} \over (z-z_i) } +{\rm finite} \; ,
\equn
$$
then 
$$
{\rm Residue}( { f(z) \over z }, z_i)  =  -{a_{-2} \over z_i^2 } +{a_{-1} \over z_i} 
\; .
\equn
$$
However for loop amplitudes only the leading singularities have been determined in general and there are no
general theorems for the sub-leading terms. We overcome this barrier by using axial gauge techniques to determine the extra information 
required to perform recursion. This is termed {\it augmented recursion}.

There are two contributions to the  factorisation:
\begin{align}
A_3^{\rm tree} \times {1\over K^2} \times A_4^{2\, \rm loop} \;\;\; {\rm and}\;\;\;  A_3^{1\, \rm loop} \times {1\over K^2} \times A_4^{1 \,\rm loop} \; .
\end{align}
The full rational term is the sum of contributions from these two channels,
\begin{align}
 R_5^{(2)}=R_5^{\rm t-2} +R_5^{1-1}\; .
\end{align}
$R_5^{\rm t-2}$ involves only single poles and is directly evaluated using the rational part of the four-point two-loop amplitude~\cite{Bern:2002tk},
\begin{align}
R_4^{(2)}(K^+,b^+,c^+,d^+)= {i\over 6 } {\spb{K}.b\over \spa{K}.b } 
{\spb{c}.d \over \spa{c}.d}\Bigl( {s_{bd}^2\over s_{cd} s_{bc}} +8\Bigr)
\; . 
\end{align}
Setting $\eta=b$, the shift excites this factorisation channel three times, giving 
\begin{align}
R_5^{\rm t-2} = \;\;&\Bigl[A^t_3(c^+,d^+,K^-){1\over s_{cd}}R_4^{(2)}(K^+,\hat e^+,a^+,b^+) \Bigr]\Bigr\vert_{\spa{\hat c}.{\hat d}=0}
\notag \\ 
+ & \Bigl[A^t_3(d^+,e^+,K^-){1\over s_{de}}R_4^{(2)}(K^+,a^+,b^+,\hat c^+) \Bigr]\Bigr\vert_{\spa{\hat d}.{\hat e}=0}
\notag \\ 
+ & \Bigl[A^t_3(e^+,a^+,K^-){1\over s_{ea}}R_4^{(2)}(K^+,b^+,\hat c^+,\hat d^+) \Bigr]\Bigr\vert_{\spa{\hat e}.{a}=0}
\; . 
\end{align}

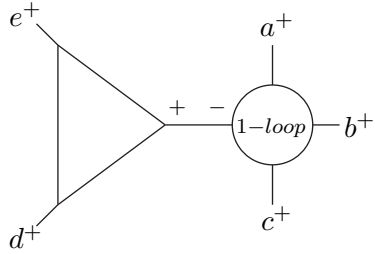
\begin{figure}[H]
\centerline{
    \begin{picture}(170,150)(-50,-20)    
     \Line( 0, 0)( 0,60)
     \Line( 0,60)(40,30)
     \Line(40,30)( 0, 0)
     \Line( 0, 0)(-8,-8)
     \Line( 0,60)(-8,68)
     \Line(40,30)(105,30)
     \Line(80,60)(80,0)
     \Text(-12,-12)[c]{$d^+$}  
     \Text(-12, 72)[c]{$e^+$}  
     \Text( 82, 68)[c]{$a^+$}  
     \Text( 82, -5)[c]{$c^+$}  
     \Text( 113, 30)[c]{$b^+$} 
     \Text( 70, 30)[c]{$\bullet$}
     \BCirc(80,30){15}
     \Text( 80, 28)[c]{${}^{1-loop}$}  
     \Text( 45, 35)[c]{$^+$}
     \Text( 60, 35)[c]{$^-$}     
     \end{picture} 
    }
    \caption{The origin of the double poles in $s_{de}$. The diagram has an explicit pole and an additional pole can arise from the triangle integral. }
    \label{doublepole}
\end{figure}

\begin{figure}[H]
\centerline{
    \begin{picture}(170,150)(-50,-20)    
     \ArrowLine( 0, 0)( 0,60)
     \ArrowLine(40,30)( 0,60)
     \ArrowLine(40,30)( 0, 0)
     \Line( 0, 0)(-8,-8)
     \Line( 0,60)(-8,68)
     \Line(40,30)(55,30)
     \Line(40,45)(40,15)
     \Text(-12,-12)[c]{$d^+$}  
     \Text(-12, 72)[c]{$e^+$}  
     \Text( 42, 53)[c]{$a^+$}  
     \Text( 42, 10)[c]{$c^+$}  
     \Text( 63, 30)[c]{$b^+$} 
     \Text( 40, 30)[c]{$\bullet$}
     \BCirc(40,30){8}
     \Text( 40, 30)[c]{$\tau^1$}  
     \Text(22,53)[c]{$\alpha^+$}
     \Text(22, 7)[c]{$\beta^-$}     
     \Text(-8,30)[c]{$\ell$}     
     \end{picture} 
    }
    \caption{The non-factorising contribution to the pole. We must also include the case with the helicities on $\alpha$ and $\beta$ reversed.}
    \label{taudiag}
\end{figure}
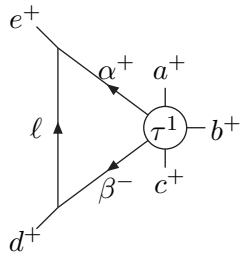

The second channel, $R^{1-1}_5$, 
has double poles associated with the diagram shown in \figref{doublepole}. The existence of double poles means we must 
determine the sub-leading contributions which are not captured by the naive factorisation. 
These {\it pole under the pole} 
contributions have been determined for a number of one-loop amplitudes using 
augmented recursion~\cite{Dunbar:2010xk,Alston:2012xd,Alston:2015gea,Dunbar:2016dgg}.  
The contribution from this channel can be 
computed using axial gauge techniques~\cite{Kosower:1989xy,Schwinn:2005pi,Vaman:2008rr}
by considering diagrams of the form shown 
in \figref{taudiag}, where $\tau^1$ represents an approximation to the doubly massive current.
A key feature of the 
axial gauge is that the internal legs have helicity assignments and vertices only involve nullified momenta as defined in eq.(\ref{nullified:eq}). 
Using the axial gauge three-point vertices, the contribution from \figref{taudiag} with the indicated helicity assignment  is
\begin{align}
C^{\alpha^+\beta^-}=\int {d^d\ell\over \ell^2\alpha^2\beta^2}  {\spa{\alpha}.b^2\over \spa{\beta}.b^2}
{[d|\ell|b\ra[e|\ell|b\ra\over \spa{d}.b\spa{e}.b} \tau^{1}(\beta^-,\alpha^+,a^+,b^+,c^+)\; ,
\label{NFintA}
\end{align}
where $\alpha$ and $\beta$ are the momenta
\begin{equation}
\beta = \ell+d \;\;\;{\rm and }\;\;\;  \alpha=-\ell+e  \; . 
\end{equation}
Within $\tau$, $\beta$ and $\alpha$ are loop-momenta dependent however the combination
$\beta+\alpha$ is not. 

As discussed in \cite{Dunbar:2016dgg}, $\tau^{1}$ does not need to capture the full off-shell behaviour of the current, but it must satisfy two conditions: 
it must reproduce the leading singularity 
as $s_{\alpha\beta}\to 0$ with $\alpha^2,\beta^2\neq 0$ (C1)
and it must reproduce the amplitude in the limit $\alpha^2,\beta^2\to 0$, $s_{\alpha\beta}\neq 0$ (C2). 
The current, as detailed in appendix~\ref{AppendixAA}, is
\begin{align}
 \tau^{1}(\beta^-,&\alpha^+,a^+,b^+,c^+)= 
 {i\over 3 } {1\over \spa{a}.b^2}
 \notag \\
 \times
 \Biggl[ &-{[\alpha c]^2[q c]\over [c\beta]\spb{\beta}.q} 
         +{\cal F}
         +[c|q|\beta\ra{\bigl([c\alpha]\spb{\alpha}.q\spb{k}.q + \spb{\alpha}.q^2\spb{c}.k\bigr)\over\spb{\beta}.q\spb{k}.q 2k.q}
 \notag \\ 
         &+{[bc]\spa{a}.c\over\spa{b}.c^2} 
         { \spa{\beta}.b^2 \over  \spa{\alpha}.b^2} \biggl(
{ \la b|\beta\alpha|b\ra  \over s_{\beta\alpha} } {[\ceta|\beta+\alpha|b\ra\over [\ceta|\beta+\alpha|a\ra}
\biggr)
\notag \\
+
\biggl( &
{[bc]\spa{a}.c\over\spa{b}.c^2} 
         { \spa{\beta}.b^2 \over  \spa{\alpha}.b^2}
{ \spa{b}.\alpha \spa{b}.a  \over \spa{\alpha}.a  } {[\ceta|\beta|b\ra\over [\ceta|\beta+\alpha|a\ra}       
         -{\spa{\beta}.a^3[a\alpha]\spa{b}.\alpha\over\spa{\beta}.c\spa{c}.b\spa{a}.\alpha^2}  \biggr)  \Biggr]\,.
\label{tau1l}
\end{align}

$C^{\alpha^+\beta^-}$ is split up into five pieces: ${\rm sl}$, ${\rm sf}$, ${\rm sk}$, ${\rm dp}$ and ${\rm ap}$ corresponding to the terms 
in $\tau$ given in \eqref{tau1l},
\begin{equation}
C^{\alpha^+\beta^-}=C^{\alpha^+\beta^-: {\rm sl}}+C^{\alpha^+\beta^-:{\rm sf} }+C^{\alpha^+\beta^-:{\rm sk}}
+C^{\alpha^+\beta^-:{\rm dp}}+C^{\alpha^+\beta^-:{\rm ap}}\; .
\end{equation}
The term $C^{\alpha^+\beta^-:{\rm dp}}$ contains the double pole and is 
\begin{align}
C^{\alpha^+\beta^-:{\rm dp}}=&\int {d^d\ell\over \ell^2\alpha^2\beta^2}  
{[d|\ell|b\ra[e|\ell|b\ra\over \spa{d}.b\spa{e}.b} 
{i\over 3} {1\over \spa{a}.b^2} {[bc]\spa{a}.{c}\over\spa{b}.c^2} { \la b|\beta\alpha|b\ra  \over s_{\beta\alpha} } 
{[\ceta|\beta+\alpha|b\ra\over [\ceta|\beta+\alpha|a\ra}
\notag \\
=&{i\over 9}{[bc]\spa{a}.{c} \la b|de|b\ra\over  \spa{a}.b^2\spa{b}.c^2  }{1\over\spa{d}.{e}^2} {[\ceta|d+e|b\ra\over [\ceta|d+e|a\ra} \; .
\label{NFintB}
\end{align}
The final term  does not contain $\spb{\beta}.q$ and is labelled $C^{\alpha^+\beta^-:{\rm ap}}$:
\begin{align}
C^{\alpha^+\beta^-:{\rm ap}}=&\int {d^d\ell\over \ell^2\alpha^2\beta^2}  
{[d|\ell|b\ra[e|\ell|b\ra\over \spa{d}.b\spa{e}.b} 
{i\over 3} {1\over \spa{a}.b^2} 
\biggl(
{[bc]\spa{a}.{c}\over\spa{b}.c^2} { \spa{b}.\alpha \spa{b}.a  \over \spa{\alpha}.a  } {[\ceta|\beta|b\ra\over [\ceta|\beta+\alpha|a\ra}
+{\spa{\beta}.a[a|\alpha|b\ra\over\spa{\beta}.{c}\spa{c}.b}\biggr)\,.
\end{align}
As this term contains only a single pole, the approximation
\begin{equation}
{\spa{X}.\alpha\over\spa{Y}.\alpha}= {\spa{X}.\alpha\over\spa{Y}.\alpha}{\spa{Y}.{d}\over\spa{Y}.{d}} = {\spa{X}.{d}\over\spa{Y}.{d}} 
+{\cal O}\bigl(\spa{\alpha}.{d}\bigr)
\end{equation}
can be used to leading order, leaving cubic triangle integrals:
\begin{align}
C^{\alpha^+\beta^-:{\rm ap}}=&  
{i\over 9} {\spb{d}.e\over \spa{d}.e}{1\over \spa{a}.b^2}
\biggl(
{[bc]\spa{a}.{c}\over\spa{b}.c^2} { \spa{b}.{d} \spa{b}.a  \over \spa{d}.a  } {[\ceta|2d+e|b\ra\over [\ceta|d+e|a\ra}
+{\spa{d}.a[a|d+2e|b\ra\over\spa{d}.{c}\spa{c}.b}\biggr) \; .
\end{align}
The term $C^{\alpha^+\beta^-:{\rm sl}}$ is
\begin{align}
C^{\alpha^+\beta^-:{\rm sl}}=&\int {d^d\ell\over \ell^2\alpha^2\beta^2}  
{[d|\ell|b\ra[e|\ell|b\ra\over \spa{d}.b\spa{e}.b} {\spa{\alpha}.b^2\over \spa{\beta}.b^2}
{i \over 3} {1\over \spa{a}.b^2}\bigl( -{[\alpha c]^2 [qc]\over [\beta q][c\beta]} \bigr)
\notag \\
=& {i\over 3} {1\over \spa{a}.b^2}{ [qc]  \over \spa{d}.b\spa{e}.b} 
\sum_{n=0,2}\int {d^d\ell\over \ell^2\alpha^2\beta^2}  
[d|\ell|b\ra[e|\ell|b\ra
{[c|\beta|b\ra^{1-n}[c|P_{de}|b\ra^n {\kappa_n}\over (\beta+q)^2} +\cdots 
\end{align}
where $\kappa_2=\kappa_0=1$, $\kappa_1=-2$ and the $+\cdots$ reflects the use of a leading order approximation based on
\begin{align}
{1\over 2\beta\cdot X}-{1\over (\beta+X)^2}={\beta^2\over 2\beta \cdot X  (\beta+X)^2} \; .
\end{align}
For $n=0,1$ this is readily reduced to triangles using
\begin{align}
 [e|\ell|b\ra[d|\ell|b\ra 
= & {\beta^2 \la b|\ell e |b\ra  + \alpha^2 \la b|\ell d|b\ra -\ell^2 \la b|(\ell-e) P_{de}|b\ra   \over \spa{e}.d} \; .
 \end{align}
As all of the numerator factors have $\ell$ contracted with $b$, only the scalar part of the shifted Feynman parameter integral survives. This
removes two of the triangles completely. Quadratic numerators in the surviving triangle give rational contributions, while linear numerators do not. 
As the $n=2$ case involves a linear box, rational contributions are not expected from this term. Overall,
\begin{align}
C^{\alpha^+\beta^-:{\rm sl}}= {i\over3 } {1\over \spa{a}.b^2}{ [qc] [de]\over \spa{d}.e} 
{[c|e|b\ra\over 2 s_{eq}}
\; . 
\end{align}

The third term in \eqref{tau1l} involves the terms with a $k^2/s_{\alpha\beta}$  factor from ${\cal F}$. These give the $C^{\alpha^+\beta^-:{\rm sk}}$ 
contribution:
\begin{align}
C^{\alpha^+\beta^-:{\rm sk}}=&{i\over 3}\int {d^d\ell\over \ell^2\alpha^2\beta^2} 
{[d|\ell|b\ra[e|\ell|b\ra\over \spa{d}.b\spa{e}.b} {\spa{\alpha}.b^2\over\spa{\beta}.b^2}
[c|q|\beta\ra{\bigl([c\alpha]\spb{\alpha}.q\spb{k}.q + \spb{\alpha}.q^2\spb{c}.k\bigr)\over\spb{\beta}.q\spb{k}.q 2k\cdot q}{1\over \spa{a}.b^2} \; .
\end{align}
Using the same leading order approximations as in the previous case, 
\begin{align}
C^{\alpha^+\beta^-:{\rm sk}}
=&-{i\over 18}\Biggl[{[cq][ed]\over \spa{a}.b^2\spa{e}.d 2k \cdot q} {1\over s_{eq}}
\Bigl[5[q|e|b\ra[c|e|b\ra + 3[q|d|b\ra[c|e|b\ra + [q|e|b\ra[c|d|b\ra  \Bigr]
\notag \\  & \hskip 3.0truecm +
{[cq][ed] [c|k|b\ra\over \spa{a}.b^2\spa{e}.d (2k\cdot q)^2} 
\Bigl[5[q|e|b\ra  +4 [q|d|b\ra\Bigr]\Biggr]\; .
\end{align}
Finally there is the contribution from the second term in \eqref{tau1l}. This term reproduces the factorising contribution shown in the second 
part of \figref{tausingbits}. The corresponding integral
\begin{align}
C^{\alpha^+\beta^-:{\rm sf}}=&\int {d^d\ell\over \ell^2\alpha^2\beta^2}  
{[d|\ell|b\ra[e|\ell|b\ra\over \spa{d}.b\spa{e}.b} {\spa{\alpha}.b^2\over \spa{\beta}.b^2}
{ \la\beta k\ra\spb{\alpha}.q^2\over \spb{\beta}.q\spb{k}.q }{1\over s_{\alpha\beta}}A^{(1)}(k^+,a^+,b^+,c^+)
\notag \\
&=C^{-+:{\rm tri}}\times {1\over s_{\alpha\beta}}A^{(1)}(k^+,a^+,b^+,c^+) \; ,
\end{align}
where the triangle integral,
\begin{align}
C^{-+:{\rm tri}}=&\int {d^d\ell\over \ell^2\alpha^2\beta^2}  
{[d|\ell|b\ra[e|\ell|b\ra\over \spa{d}.b\spa{e}.b} {\spa{\alpha}.b^2\over \spa{\beta}.b^2}
{ \la\beta k\ra\spb{\alpha}.q^2\over \spb{\beta}.q\spb{k}.q } \; ,
\end{align}
is closely related to  the $(+,+,-)$ one-loop splitting function.
\def\TRIbit{

To extract the rational part of the integral consider a 4-dimensional parametrisation
of the loop momentum integral:
\begin{align}
 \ell \to A \lambda_d\bar\lambda_d +B \lambda_e\bar\lambda_e  +     C{\spb{q}.e\over \spb{q}.d} \lambda_e\bar\lambda_d
                                                            +\bar C{\spb{q}.d\over \spb{q}.e} \lambda_d\bar\lambda_e\,,
\end{align}
so that,
\begin{align}
C^{-+:{\rm tri}} \to&{[qd][qe][ed]\over \spb{k}.q^2}
\int {dAdBdCd\bar C}{1\over AB-C\bar C} {1\over (A+1)B-C\bar C} {1\over A(B-1)-C\bar C}  
\notag \\ & \times
{\bigl( B +\bar C \chi\bigr)
 \bigl( A +     {C\over \chi}\bigr)}
{ \Bigl( \bigl(A+\bar C\bigr)\chi +\bigl( (B-1)+ C\bigr) \Bigr)^2
                                                            \over
 \Bigl( \bigl(A+1+\bar C\bigr)\chi +\bigl( B+ C\bigr) \Bigr)^2
}
\Bigl( \bigl(A +\bar C\bigr) -\bigl( (B-1)+C\bigr) \Bigr)
\end{align}
where $\chi={[q|d|b\ra\over [q|e|b\ra}$.  Under the reparameterisation  $A \leftrightarrow -B$, $C \leftrightarrow -\bar C$,
\begin{align}
C^{-+:{\rm tri}}(\chi)\to C^{+-:{\rm tri}}(1/\chi)
\end{align}
hence the sum of the two can be written as,
\begin{align}
C^{-+:{\rm tri}}+C^{+-:{\rm tri}}={[qd][qe][ed]\over \spb{k}.q^2}\times f(\chi)
\end{align}
where $f(\chi)=f(1/\chi)$. This also ensures the overall antisymmetry of the triangle under flipping, i.e. $d \leftrightarrow e$.

In the real collinear limit $d\to zk$, $e\to (1-z)k$,
\begin{align}
{1\over s_{de}}\Bigl(C^{-+:{\rm tri}}+C^{+-:{\rm tri}}\Bigr)
=&\sqrt{z(1-z)}{1\over \spa{d}.e} \times f\bigl({z\over 1-z}\bigr)
\end{align}
Comparing with the relevant term in the one-loop splitting function,
\begin{align}
f\bigl({z\over 1-z}\bigr) ={1\over 3} 
\end{align}
}   
Comparing with the one-loop splitting function leads to 
\begin{align}
C^{\alpha^+\beta^-:{\rm sf}}+C^{\alpha^+\beta^-:{\rm sf}}={1\over 3}{[qd][qe][ed]\over \spb{k}.q^2}\times {1\over s_{\alpha\beta}}A^{(1)}(k^+,a^+,b^+,c^+) \; .
\end{align}

Having determined the rational contributions arising from \figref{taudiag} the corresponding residues can be obtained by applying the shift~\eqref{KasperShift}
and extracting the coefficient  of the $(z-z_0)^{-1}$ term in the Laurent expansion. The process can be repeated for the other internal helicity configuration 
of the triangle. A similar procedure can be applied to the other two factorisation channels: $\spa{\hat c}.{\hat d}\to 0$ and $\spa{\hat e}.{a}\to 0$. As 
$\lambda_q=\lambda_b$ the five-point single-minus amplitudes in these cases need to written in a form where the terms containing 
the $\spa{\alpha}.\beta\to 0$ pole reproduce the axial gauge factorisation.

Summing over the various contributions yields a rational term that has the correct cyclic symmetry and is independent of $\bar\lambda_q$.  
These are highly non-trivial checks since these symmetries are not manifest during the recursive calculation and are only restored at the final stage 
(provided all terms have been
correctly computed).

After some considerable algebra, these terms can be reduced to match the form 
given in ref.~\cite{Gehrmann:2015bfy}~\footnote{We find a perfect match provided we replace $\tr_{-}$ of 
ref.~\cite{Gehrmann:2015bfy} by $\tr_{+}$ in term $R_5^b$.  The $\tr_{+}$ of $R_5^b$ correctly gives the collinear limit as demonstrated in 
appendix~\ref{AppendixBB}.}
\begin{equation}
R_5^{(2)}=  {  i \over 6 \spa1.2\spa2.3\spa3.4\spa4.5\spa5.1 } \times \left( R_5^a +R_5^b \right)\; ,
\end{equation}
where
\begin{align}
R_5^a &= \frac{2}{3} \sum { \tr_+^2 (4512) \over s_{45}s_{12} } \; ,
\notag \\
R_5^b &= \sum \left( \frac{10}{3} s_{12}s_{23} +\frac{2}{3} s_{12}s_{34}  \right) 
\label{ratterms:eq}
 \end{align}
and the sum cycles the five indices.

\section{Conclusions} 

Using four dimensional unitarity and recursion we have been able to reproduce the two-loop five-point all plus Yang-Mills amplitude.  
Key to this is the observation that four dimensional unitarity can be used to generate the IR singular terms whose coefficient, the one-loop amplitude, 
can be promoted to its all-$\epsilon$ form.  With this identification the finite remainder terms follow.  Computation of the cut-constructible terms is 
straightforward while computing the rational terms is fairly complicated but only involves 
one-loop integrals and avoids genuine two-loop integration.  We intend to apply these techniques to further ``pseudo-one-loop'' amplitudes~\cite{DJP}.

\section{Acknowledgements}

This work was supported by STFC grant ST/L000369/1.

\appendix

\section{Off-Shell Current}
\label{AppendixAA}

In this appendix we compute an effective current $\tau^1(\alpha^+,\beta^-,c^+,d^+,e^+)$ where $\alpha$ and $\beta$ are the off-shell legs. We will not generate the 
exact current but one which is sufficient to determine the poles in the amplitude. Specifically, as shown in~\cite{Dunbar:2016dgg}, $\tau^{1}$ must satisfy
two conditions:  (C1)
it must reproduce the leading singularity 
as $s_{\alpha\beta}\to 0$ with $\alpha^2,\beta^2\neq 0$
and (C2) it must reproduce the amplitude in the limit $\alpha^2,\beta^2\to 0$, $s_{\alpha\beta}\neq 0$. 

We use an axial gauge formalism~\cite{Kosower:1989xy,Schwinn:2005pi,Vaman:2008rr} in which helicity labels can be used for internal lines and 
off-shell internal legs in the vertices are nullified using a reference spinor: given a reference null momentum $\eta$, any off-shell leg with momentum $K$ 
can be {\it nullified} using 
$$
K^\flat= K- {K^2\over [\eta| K | \eta \rangle }  \eta \; ,
\equn$$
which gives spinors
$$
\lambda_K= \alpha K|\eta]  \;  , \;\   \bar\lambda_K= \alpha^{-1} { K|\eta\rangle \over [\eta| K | \eta \rangle } \; .
\equn
\label{nullified:eq}
$$  
For convenience we will choose 
the reference spinor to be $q=\bar\lambda_q \lambda_b$ leaving $\bar\lambda_q$ arbitrary.

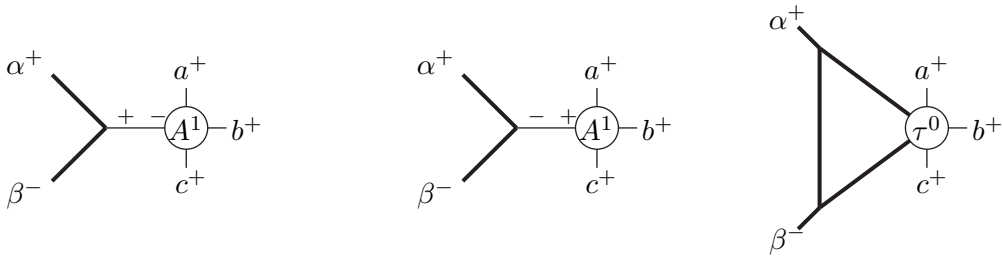
\begin{figure}[H]
\centerline{
    \begin{picture}(150,150)(-30,-20)  
     \SetWidth{1.5}
     \Line(20,50)(40,30)
     \Line(20,10)(40,30)    
     \SetWidth{0.5}
     \Line(40,30)(85,30)
     \Line(70,45)(70,15)
     \Text( 10, 5)[c]{$\beta^-$}  
     \Text( 10, 55)[c]{$\alpha^+$}  
     \Text( 72, 53)[c]{$a^+$}  
     \Text( 72, 10)[c]{$c^+$}  
     \Text( 93, 30)[c]{$b^+$} 
     \Text( 70, 30)[c]{$\bullet$}
     \Text( 48, 33)[c]{$^+$}
     \Text( 60, 33)[c]{$^-$}     
     \BCirc(70,30){8}
     \Text( 70, 30)[c]{$A^1$}  
     \end{picture}  
     \begin{picture}(150,150)(-30,-20)  
     \SetWidth{1.5}
     \Line(20,50)(40,30)
     \Line(20,10)(40,30)    
     \SetWidth{0.5}
     \Line(40,30)(85,30)
     \Line(70,45)(70,15)
     \Text( 10, 5)[c]{$\beta^-$}  
     \Text( 10, 55)[c]{$\alpha^+$}  
     \Text( 72, 53)[c]{$a^+$}  
     \Text( 72, 10)[c]{$c^+$}  
     \Text( 93, 30)[c]{$b^+$} 
     \Text( 70, 30)[c]{$\bullet$}
     \Text( 48, 33)[c]{$^-$}
     \Text( 60, 33)[c]{$^+$}     
     \BCirc(70,30){8}
     \Text( 70, 30)[c]{$A^1$}  
     \end{picture}  
     \begin{picture}(150,150)(-30,-20) 
     \SetWidth{1.5}
     \Line( 0, 0)( 0,60)
     \Line(40,30)( 0,60)
     \Line(40,30)( 0, 0)
     \Line( 0, 0)(-8,-8)
     \Line( 0,60)(-8,68)
     \SetWidth{0.5}
     \Line(40,30)(55,30)
     \Line(40,45)(40,15)
     \Text(-12,-12)[c]{$\beta^-$}  
     \Text(-12, 72)[c]{$\alpha^+$}  
     \Text( 42, 53)[c]{$a^+$}  
     \Text( 42, 10)[c]{$c^+$}  
     \Text( 63, 30)[c]{$b^+$} 
     \Text( 40, 30)[c]{$\bullet$}
     \BCirc(40,30){8}
     \Text( 40, 30)[c]{$\tau^0$}  
     \end{picture} 
    }
    \caption{Sources of $s_{\alpha\beta}$ poles in $\tau^1$}
    \label{tausingbits}
\end{figure}

Our task is to identify the part of the current which will generate $s_{\alpha\beta}^{-1}$ poles. The
diagrams which lead to these poles are shown in fig.~\ref{tausingbits}. 

The first diagram of fig.\ref{tausingbits} contains a $\spa{\alpha}.\beta^{-1}$ factor and hence, after the integration within 
the diagram as in fig.\ref{taudiag} generates the double pole piece of the rational terms. The second diagram contains
a $\spb{\alpha}.\beta^{-1}$ factor
and so does not enhance the order of the $\spa{d}.e$ pole.

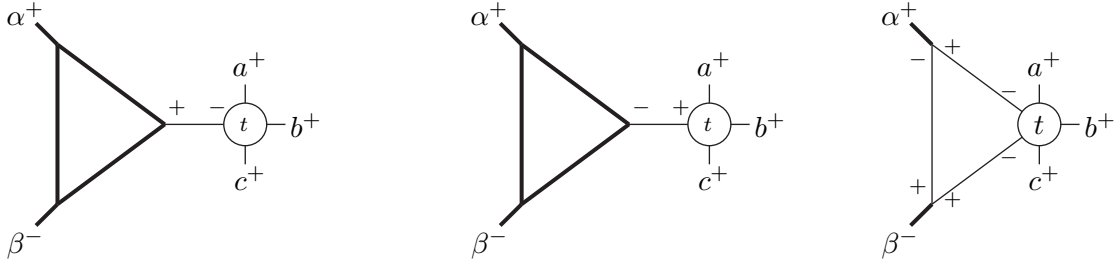
\begin{figure}[H]
\centerline{
    \begin{picture}(170,150)(-50,-20)    
     \SetWidth{1.5}
     \Line( 0, 0)( 0,60)
     \Line( 0,60)(40,30)
     \Line(40,30)( 0, 0)
     \Line( 0, 0)(-8,-8)
     \Line( 0,60)(-8,68)
     \SetWidth{0.5}
     \Line(40,30)(85,30)
     \Line(70,45)(70,15)
     \Text(-12,-15)[c]{$\beta^-$}  
     \Text(-12, 72)[c]{$\alpha^+$}  
     \Text( 72, 53)[c]{$a^+$}  
     \Text( 72, 10)[c]{$c^+$}  
     \Text( 93, 30)[c]{$b^+$} 
     \Text( 70, 30)[c]{$\bullet$}
     \BCirc(70,30){8}
     \Text( 70, 28)[c]{$^{t}$}  
     \Text( 45, 35)[c]{$^+$}
     \Text( 60, 35)[c]{$^-$}     
     \end{picture} 
     \begin{picture}(170,150)(-50,-20)    
     \SetWidth{1.5}
     \Line( 0, 0)( 0,60)
     \Line( 0,60)(40,30)
     \Line(40,30)( 0, 0)
     \Line( 0, 0)(-8,-8)
     \Line( 0,60)(-8,68)
     \SetWidth{0.5}
     \Line(40,30)(85,30)
     \Line(70,45)(70,15)
     \Text(-12,-15)[c]{$\beta^-$}  
     \Text(-12, 72)[c]{$\alpha^+$}  
     \Text( 72, 53)[c]{$a^+$}  
     \Text( 72, 10)[c]{$c^+$}  
     \Text( 93, 30)[c]{$b^+$} 
     \Text( 70, 30)[c]{$\bullet$}
     \BCirc(70,30){8}
     \Text( 70, 28)[c]{$^{t}$}  
     \Text( 45, 35)[c]{$^-$}
     \Text( 60, 35)[c]{$^+$}     
     \end{picture} 
     \begin{picture}(150,150)(-30,-20) 
      \Line( 0, 0)( 0,60)
     \Line(40,30)( 0,60)
     \Line(40,30)( 0, 0)
     \SetWidth{1.5}    
     \Line( 0, 0)(-8,-8)
     \Line( 0,60)(-8,68)
     \SetWidth{0.5}
     \Line(40,30)(55,30)
     \Line(40,45)(40,15)
     \Text(-12,-15)[c]{$\beta^-$}  
     \Text(-12, 72)[c]{$\alpha^+$}  
     \Text( 42, 53)[c]{$a^+$}  
     \Text( 42, 10)[c]{$c^+$}  
     \Text( 63, 30)[c]{$b^+$} 
     \Text( 40, 30)[c]{$\bullet$}
     \BCirc(40,30){8}
     \Text( 40, 30)[c]{$t$}  
     \Text(29,41)[c]{$^-$}
     \Text(29,16)[c]{$^-$}
     \Text( 8,58)[c]{$^+$}
     \Text( 8,0)[c]{$^+$}
     \Text( -5,53)[c]{$^-$}
     \Text( -5,5)[c]{$^+$}
     \end{picture} 
    }
    \caption{Sources of $s_{\alpha\beta}$ poles in contributions from $\tau^0$}
    \label{tautreebits}
\end{figure}

The possible sources of $s_{\alpha\beta}$ poles in the third structure are illustrated in \figref{tautreebits}. With this helicity configuration
the $({\rm triangle})\times({\rm tree})$ factorisations with
$\beta^-$ in the triangle are absent as there are insufficient negative helicity legs to form a non-vanishing tree. 
Also, any triangles involving $\beta^-$ and $\alpha^+$ must be {\it mixed} (i.e. contain both $(++-)$ and $(--+)$ corners) and are therefore finite. 
This removes contributions of the form 
$({\rm singular\; triangle })\times({\rm on-shell\; propagator})\times ({\rm current \;with\;vanishing \;amplitude})$.
As there are no contributions with a $1/s_{\alpha\beta}$ propagator, any poles in $s_{\alpha\beta}$ must comes from the loop 
integration. Such singularities arise from the integration region with the loop momenta all proportional to $\alpha+\beta$, i.e. a specific null momentum. For
these contributions the loop momenta can be taken to be on-shell (hence the of thin lines for the propagators in the third part of \figref{tautreebits}).
While there is a helicity configuration which gives a non-vanishing tree amplitude for the third corner, this amplitude vanishes when the propagators are
collinear, i.e. the tree vanishes in the region of interest and the contribution is finite as $s_{\alpha\beta}\to 0$. Thus there are no poles in $s_{\alpha\beta}$
arising from the third structure in \figref{tausingbits} and it can be neglected when considering condition C1 (the finite contributions of course are relevant
for condition C2).


$\tau^1$ can be constructed from the five-point one-loop amplitude~\cite{Bern:1993mq}:
\begin{equation}
 A_5^{(1)}(\beta^-,\alpha^+,a^+,b^+,c^+)= {i\over 3} {1\over \spa{a}.b^2}
 \Biggl[ -{[\alpha c]^3\over [\beta\alpha][c\beta]} 
         +{\spa{\beta}.b^3[bc]\spa{a}.c\over\spa{\beta}.\alpha\spa{\alpha}.a\spa{b}.c^2} 
         -{\spa{\beta}.a^3[a\alpha]\spa{b}.\alpha\over\spa{\beta}.c\spa{c}.b\spa{a}.\alpha^2}  \Biggr]\,.
 \label{spec5pt}
 \end{equation}
To satisfy C1 without compromising C2, corrections of order $\alpha^2$ and $\beta^2$ are introduced to reproduce the factorisation channels in \figref{tausingbits}.
Using axial gauge rules and the one-loop amplitude~\cite{Bern:2002tk}
\begin{align}
 A_4^{(1)}(d^-,a^+,b^+,c^+)=-{i\over 3} {\spb{a}.c^2 s_{ac}\over \spb{d}.a\spa{a}.b\spa{b}.c\spb{c}.d} \; ,
\end{align}
the pole arising from the first structure is
\begin{align}
&{ [\alpha k]\spa{\beta}.q^2\over \spa{\alpha}.q\spa{k}.q }{1\over s_{\alpha\beta}}A^{(1)}(k^-,a^+,b^+,c^+)
\notag \\
=&-{i\over 3} {\spa{\beta}.q^2\over \spa{\alpha}.q^2}
{ \la q|\alpha \beta|q\ra\over s_{\alpha\beta}}{\spb{a}.c^2 s_{ac}\over \spb{k}.a\spa{a}.b\spa{b}.c\spb{c}.k \spa{k}.q^2} \; ,
\end{align}
where $k=-k_a-k_b-k_c$ which is null on the pole. 
With $\lambda_q\to \lambda_b$ the four-point kinematics on the loop amplitude allow this to be written as
\begin{align}
&-{i\over 3} {\spa{\beta}.b^2\over \spa{\alpha}.b^2}
{ \la b|\alpha \beta|b\ra\over s_{\alpha\beta}}{\spb{a}.c^2 s_{ac}\over \spb{k}.a\spa{a}.b\spa{b}.c\spb{c}.k \spa{k}.b^2}
\notag \\
=&-{i\over 3} {\spa{\beta}.b^2\over \spa{\alpha}.b^2}
{ \la b|\alpha \beta|b\ra\over s_{\alpha\beta}}{\spa{a}.c\spb{b}.c \spa{k}.b\over \spa{a}.b^2\spa{b}.c^2 \spa{k}.a} \; .
\label{targApole}
\end{align}
This factor can be built into  $\tau^1$ by taking \eqref{spec5pt} and making the substitution
 \begin{align}
{ \spa{\beta}.b^3 \over  \spa{\beta}.\alpha \spa{\alpha}.a }
\to & { \spa{\beta}.b^2 \over  \spa{\alpha}.b^2} \biggl(
{ \la b|\beta\alpha|b\ra  \over s_{\beta\alpha} } {[\ceta|\beta+\alpha|b\ra\over [\ceta|\beta+\alpha|a\ra}
+
{ \spa{b}.\alpha \spa{b}.a  \over \spa{\alpha}.a  } {[\ceta|\beta|b\ra\over [\ceta|\beta+\alpha|a\ra} \biggr)
\label{magicsubA}
\end{align}
in the second term.
\eqref{magicsubA} is an identity in the limit $\alpha^2,\beta^2\to 0$ and so condition C2 is not compromised. 
The leading term as  $s_{\alpha\beta}\to 0$ then  exactly reproduces the contribution from \eqref{targApole}.

Similarly the second structure gives
\begin{align}
{\cal F}=&{ \la\beta k\ra\spb{\alpha}.q^2\over \spb{\beta}.q\spb{k}.q }{1\over s_{\alpha\beta}}A^{1-l}(k^+,a^+,b^+,c^+)
=-{i\over 3}{ \la\beta k\ra\spb{\alpha}.q^2\over \spb{\beta}.q\spb{k}.q }{1\over s_{\alpha\beta}}{\spb{c}.k^2\over\spa{a}.b^2} \; .
\end{align}
Away from the pole $k$ is interpreted as its nullified form so that

\begin{align}
{\cal F} &={i\over 3}{\spb{\alpha}.q^2\over \spb{\beta}.q\spb{k}.q }{1\over s_{\alpha\beta}}{\spb{c}.k 
\bigl([c\alpha]\spa{\alpha}.\beta +\delta[c|q|\beta\ra\bigr)\over\spa{a}.b^2}
\notag \\
&=
{i\over 3}{1\over s_{\alpha\beta}}
{\bigl(\spb{\alpha}.q\spb{k}.q\spb{c}.k[c\alpha]\spa{k}.\beta +\delta[c|q|\beta\ra \spb{\alpha}.q^2\spb{c}.k\bigr)\over\spb{\beta}.q\spb{k}.q\spa{a}.b^2}
\notag \\
&=
{i\over 3}{1\over s_{\alpha\beta}}\Biggl[
{[c\alpha]^2\spb{\alpha}.q \spa{\alpha}.{\beta} \over\spb{\beta}.q\spa{a}.b^2}
+\delta[c|q|\beta\ra{\bigl([c\alpha]\spb{\alpha}.q\spb{k}.q + \spb{\alpha}.q^2\spb{c}.k\bigr)\over\spb{\beta}.q\spb{k}.q\spa{a}.b^2}
\Biggr] \; ,
\label{Fexpn}
\end{align}
where 
\begin{align}
\delta={\alpha^2\over 2\alpha\cdot q}+{\beta^2\over 2\beta\cdot q} - {k^2\over 2k\cdot q} \; . 
\end{align}
Now,
\begin{align}
{\spa{\alpha}.{\beta} \over s_{\alpha\beta}} -{1\over \spb{\beta}.{\alpha}}
&={\spa{\alpha}.{\beta}\spb{\beta}.{\alpha} -s_{\alpha\beta}\over s_{\alpha\beta}\spb{\beta}.{\alpha}}
={(\alpha^\flat+\beta^\flat)^2 -s_{\alpha\beta}\over s_{\alpha\beta}\spb{\beta}.{\alpha}}
=-\biggl({\alpha^2\over 2\alpha\cdot q}+{\beta^2\over 2\beta\cdot q}\biggr) {2k\cdot q\over s_{\alpha\beta}\spb{\beta}.{\alpha}}
\label{sgames}
\end{align}
and the first term in the amplitude is
\begin{align}
& -{i \over 3} {1\over \spa{a}.b^2}{[\alpha c]^3\over [\beta\alpha][c\beta]} 
=
-{i\over 3} {1\over \spa{a}.b^2}{[\alpha c]^3\over [\beta\alpha][c\beta]}{\spb{\beta}.q\over\spb{\beta}.q}
\notag \\
=&
-{i\over 3} {1\over \spa{a}.b^2}{[\alpha c]^2\over [\beta\alpha][c\beta]\spb{\beta}.q}{\spb{\beta}.q[\alpha c]}
 =
 -{i\over 3} {1\over \spa{a}.b^2}{[\alpha c]^2[q c]\over [c\beta]\spb{\beta}.q}
-{i\over 3} {1\over \spa{a}.b^2}{[\alpha c]^2[q\alpha]\over [\beta\alpha]\spb{\beta}.q} \; .
\label{Abitexpn}
 \end{align}
Using \eqref{sgames} the second term of \eqref{Abitexpn} matches the first term of \eqref{Fexpn} up to corrections of order $\alpha^2$ and $\beta^2$. 
However, in addition to terms of order $\alpha^2$ and $\beta^2$, the second term of \eqref{Fexpn} contains a term of order $k^2/s_{\alpha\beta}$. This term
does not contribute to the $s_{\alpha\beta}^{-1}$ pole in $\tau^{1}$ and is not present in the amplitude when $\alpha^2,\beta^2\to 0$. The current is 
therefore obtained 
by replacing the second term of \eqref{Abitexpn} by ${\cal F}$ with the order $k^2/s_{\alpha\beta}$ term removed. The current is then
\begin{align}
 \tau^{1}(\beta^-,&\alpha^+,a^+,b^+,c^+)= 
 {i\over 3} {1\over \spa{a}.b^2}
 \notag \\
 \times
 \Biggl[ &-{[\alpha c]^2[q c]\over [c\beta]\spb{\beta}.q} 
         +{\cal F}
         +[c|q|\beta\ra{\bigl([c\alpha]\spb{\alpha}.q\spb{k}.q + \spb{\alpha}.q^2\spb{c}.k\bigr)\over\spb{\beta}.q\spb{k}.q 2k\cdot q}
 \notag \\ 
         &+{[bc]\spa{a}.c\over\spa{b}.c^2} 
         { \spa{\beta}.b^2 \over  \spa{\alpha}.b^2} \biggl(
{ \la b|\beta\alpha|b\ra  \over s_{\beta\alpha} } {[\ceta|\beta+\alpha|b\ra\over [\ceta|\beta+\alpha|a\ra}
+
{ \spa{b}.\alpha \spa{b}.a  \over \spa{\alpha}.a  } {[\ceta|\beta|b\ra\over [\ceta|\beta+\alpha|a\ra} \biggr)
         -{\spa{\beta}.a^3[a\alpha]\spa{b}.\alpha\over\spa{\beta}.c\spa{c}.b\spa{a}.\alpha^2}  \Biggr]\,.
 \label{tau1l:app}
 \end{align}
where, by construction, the third term exactly reproduces the first structure in \figref{tausingbits} and the second term gives the $s_{\alpha\beta}^{-1}$ 
pole in the second 
structure in \figref{tausingbits}. This expression therefore satisfies condition C1. The modifications to the amplitude are all ${\cal O}(\alpha^2,\beta^2)$
and therefore do not compromise condition C2.

\section{Collinear Limits}
\label{AppendixBB}

We consider the collinear limit of the amplitude as an important consistency test and to illustrate some key features.  The collinear limit
occurs when adjacent momenta $k_a$ and $k_b$ become collinear,
\begin{equation}
k_a \longrightarrow z \times K , \;\;\ k_b \longrightarrow (1-z) \times K = \bar z K \; .
\end{equation}
In this limit amplitudes factorise as 
\begin{equation}
A^{(L)}_n (\cdots ,k_a^{h} , k_b^{h'}, \cdots )
\longrightarrow  
\sum_{L_s,h''}    S^{hh', (L_s) }_{-h''} \times A^{(L-L_s)}_{n-1} (\cdots K^{h''}, \cdots )\; ,
\end{equation}
where $S^{hh',(L_s)}_{-h''}$ are the various splitting functions.   For our amplitude the tree amplitude vanishes for both choices of $h''$ and
\begin{equation}
A^{(2)}_5 (\cdots k_a^{+} , k_b^{+} \cdots )
\longrightarrow  
 S^{++, \tree }_{-} \times A^{(2)}_{4} (\cdots K^{+}, \cdots )
 +\sum_{h''=\pm} S^{++,(1)}_{\pm}\times  A^{(1)}_{4} (\cdots K^{\mp}, \cdots ) \; .
\end{equation}
The first important result is that {\it to all orders in $\epsilon$},
\begin{equation}
A^{(1)}_5 \longrightarrow S^{++,\tree}_- \times A^{(1)}_4 \; .
\end{equation}
The all-$\epsilon$ forms of these amplitudes are~\cite{Bern:1996ja} 
\begin{align}
A^{(1)}_{4}&(1^+,2^+,3^+,4^+) =
{ 2i \eps(1-\eps) \over  \spa1.2\spa2.3\spa3.4\spa4.1}
\times s_{12}s_{23} I_4^{D=8-2\eps} \,,
\notag 
\\
A^{(1)}_{5}&(1^+,2^+,3^+,4^+,5^+) =
{ i \eps(1-\eps)\over  \spa1.2\spa2.3\spa3.4\spa4.5\spa5.1}
\notag \\
&\times
\Bigl[
s_{23}s_{34} I_4^{(1),D=8-2\eps}
+s_{34}s_{45} I_4^{(2),D=8-2\eps}
+s_{45}s_{51} I_4^{(3),D=8-2\eps}
\cr
&+s_{51}s_{12} I_4^{(4),D=8-2\eps}
+s_{12}s_{23} I_4^{(5),D=8-2\eps}
+(4-2\eps) { \varepsilon (1,2,3,4) }I_5^{D=10-2\eps}
\Bigr] \; .
\end{align}
In the collinear limit the pentagon $I_5^{D=10-2\eps}$ does not contribute since its coefficient vanishes for four-point kinematics. 
The one-mass boxes do not individually become the massless box however, by examining the hypergeometric representation of these functions~\cite{Bern:1993kr}
we see that they combine to all orders in $\epsilon$ to yield the massless box.  
This is quite important because the expansion in $\epsilon$ of the boxes, e.g. the massless box
\begin{align}
I_4^{D=8-2\epsilon}= 
 { -1 \over 2 \eps (3-2\epsilon)(1-2\eps) } \left( { st \over u^2 }\right) \biggl( { u^2 \over st }
+&\epsilon \biggl(  -{u^2 \over st }+\Log^2[s/t]/2
\biggr)
 \\
 & +\epsilon^2 \biggl( -{u^2 \over st }+\Li_3(1+s/t)+\Li_3(1+t/s)
\biggr)\; ,
\end{align}
involves more complex functions including polylogarithms. These, when multiplied by the IR singular terms contribute 
to the amplitude. 

Next, consider the IR singular factor, 
\begin{equation}
F_n^0 =  \sum_i  -{1\over \epsilon^2} \left( {-s_{ii+1} } \right)^{-\epsilon}  \; .
\end{equation}
In the collinear limit, 
\begin{equation}
F_n^0 \longrightarrow F_{n-1}^0 +r^{++}_- +\Delta \; ,
\end{equation}
where~\cite{Bern:1994zx}  
\begin{equation}
r^{++}_-
=
 - {1\over\epsilon^2}\left( {\mu^2\over z(1-z)(-s_{ab})}\right)^{\epsilon}
 + 2 \ln z\,\ln(1-z) + {1\over3} z (1-z) - {\pi^2\over6}
\end{equation}
and 
\begin{equation}
\Delta=\log(s_{ab})\log(z\bar z)-\log(s_{a-1,a})\log(z)-\log(s_{b,b+1})\log(\bar z)
-\log(z)\log(\bar z) -\frac{1}{3} z \bar z +\frac{\pi^2}{4} \; .
\end{equation}
The combination  $S^{++,\tree}_- \times r^{++}_{-}$ is the one-loop splitting function.

Consequently, 
\begin{align}
A^{(1)}_5 \times F^0_5 &\longrightarrow 
S^{++,\tree}_- A^{(1)}_4   \left(   F_4^0 + r^{++}_- +\Delta \right)
\notag \\
&=  S^{++,\tree}_- \bigl( A^{(1)}_4 F_4^0 \bigr)
+\left(S^{++,\tree}_- r^{++}_- \right) A^{(1)}_4
+S^{++,\tree}_- A^{(1)}_4 \Delta \; .
\end{align}
In the last term, $S^{++,\tree}_- A^{(1)}_4 \Delta$, we need only keep the one-loop amplitude to order $\epsilon^0$.  

When we consider the remainder function of eq.~(\ref{eq:cc}) in the collinear limit we find
\begin{equation}
F_5^{cc} \longrightarrow   - S^{++,\tree}_- A^{(1)}_4 \Delta +\hbox{\rm rational terms}\; .
\end{equation}
This is consistent with the absence of a $F^{cc}_4$ term in the four-point amplitude. 

The rational terms $R_5^{(2)}$ must satisfy  
\begin{equation}
R_5^{(2)}  \longrightarrow S^{++,\tree}_{-} \times R_4^{(2)}(++++)   +S^{++,(1)}_{+} \times A_4^{(1)}(+++-)  +S^{++,(1)}_{-}\big|_{\rm rat} \times A_4^{(1)}(++++)\; .
\end{equation}
where $S^{++,(1)}_{-}\big|_{\rm rat}$ is the rational part of the splitting function. 

$S^{++}_{+}A_4^{(1)}(+++-) $ arises as a  $\spb{a}.b/\spa{a}.b^2$ pole which is a double pole for  complex momenta.  If we consider 
$a=4,b=5$, for example, two of the terms in $R_5^a$ contribute. These terms are, (using the terms of eq.~(\ref{ratterms:eq}) rather than the form in 
ref.~\cite{Gehrmann:2015bfy})
\begin{equation}
{i \over 9} 
{ \spb1.2 \spb4.5 \over \spa1.2^2\spa4.5^2} 
{ \spa2.4^2 \spa5.1 \over \spa2.3\spa3.4 }
+
{i \over 9} 
{ \spb4.5 \spb2.3 \over \spa4.5^2\spa2.3^2} 
{ \spa5.2^2 \spa3.4 \over \spa5.1\spa1.2 } \; .
\end{equation}
The $45$ collinear limit of this is then (with some algebraic manipulation) 
\begin{align}
{i \over 9} \times
{ \sqrt{z\bar{z} }  \spb4.5 \over \spa4.5^2} 
&\times 
\left( 
{ \spb1.2  \over \spa1.2^2} 
{ \spa2.4^2 \spa4.1 \over \spa2.3\spa3.4 }
+
{  \spb2.3 \over \spa2.3^2} 
{ \spa4.2^2 \spa3.4 \over \spa4.1\spa1.2 }
\right)
\notag \\
=
{ -\sqrt{z\bar{z} }  \spb4.5 \over 3 \spa4.5^2} 
&\times 
\left(
{ - i \spb1.3^2 u 
\over 
3 \spb{K}.1 \spa1.2 \spa2.3 \spb3.{K} }
\right)
 =  S^{++,(1)}_+ 
 \times A^{(1)}_4 (+++-)
\end{align}
as required.


\end{document}